\documentclass[english]{article}
\usepackage[T1]{fontenc}
\usepackage[latin9]{inputenc}
\usepackage{geometry}
\usepackage{color}
\usepackage{array}
\usepackage{float}

\makeatletter

\providecommand{\tabularnewline}{\\}

\makeatother

\usepackage{babel}

\begin{document}

\title{Orthogonal-state-based protocols of quantum key agreement}

\author{Chitra Shukla$^{1}$, Nasir Alam$^{2}$ and Anirban Pathak$^{1,3}$}

\maketitle
\begin{center}
$^{1}$Jaypee Institute of Information Technology, A-10, Sector-62,
Noida, India
\par\end{center}

\begin{center}
$^{2}$Department of Physics Visva Bharati, Santiniketan, West Bengal-731235,
India
\par\end{center}

\begin{center}
$^{3}$RCPTM, Joint Laboratory of Optics of Palacky University and 
\par\end{center}

\begin{center}
Institute of Physics of Academy of Science of the Czech Republic,
Faculty of Science, Palacky University, 17. listopadu 12, 771 46 Olomouc,
Czech Republic
\par\end{center}
\begin{abstract}
Two orthogonal-state-based protocols of quantum key agreement (QKA)
are proposed. The first protocol of QKA proposed here is designed
for two-party QKA, whereas the second protocol is designed for multi-party
QKA. Security of these orthogonal-state-based protocols arise from
monogamy of entanglement. This is in contrast to the existing protocols
of QKA where security arises from the use of non-orthogonal state
(non-commutativity principle). Further, it is shown that all the quantum
systems that are useful for implementation of quantum dialogue and
most of the protocols of secure direct quantum communication can be
modified to implement protocols of QKA.
\end{abstract}

\section{Introduction}

Since Bennett and Brassard \cite{bb84} proposed the first protocol
of unconditionally secure quantum key distribution (QKD), several
aspects of secure quantum communication have been explored \cite{ekert,b92,vaidman-goldenberg,ping-pong,lm05,Hierarchical}.
One such idea is quantum key agreement (QKA) \cite{N. Zhou (2004),QKA based on BB84,Three party QKA,gissin}.
There are two notions of QKA. In the weaker notion of QKA that was
followed in \cite{QKA-1998} a key is generated by two or more parties
through the negotiation which happens in public. Under this weaker
notion of QKA many of the existing protocols of QKD can be viewed
as protocols of QKA. For example, well-known BB84 \cite{bb84}, Ekert
\cite{ekert} and B92 \cite{b92} protocols of QKD qualify as protocols
of QKA if we follow the weaker notion of QKA introduced in \cite{QKA-1998}.
However, we are interested in a stronger notion of QKA that was introduced
in Ref. \cite{N. Zhou (2004)} and is subsequently followed in all
the recent works on QKA \cite{QKA based on BB84,Three party QKA,C.W. Tasi,Tsai2,Improvement on QKA,MQKA,Improvements on MQKA,multi-party,MQKA-crypto analysis}.
In this notion of QKA all the parties involved in the key generation
process contribute equally to construct the key. This is in contrast
to QKD where a single party can control the entire key. Before we
introduce new protocols of QKA it is important to understand the differences
between key distribution (KD) and key agreement (KA) in further detail.
In a KD protocol, a trusted authority (TA) chooses a secret key that
will be used in future for communication, and transmits (distributes)
it to other parties who want to communicate. In contrast, in a KA
scheme (KAS): two or more parties establish a secret key on their
own. Thus \textcolor{black}{in two-party scenario we may say that
in protocols of KD, a key is created by Alice and the same is securely
transmitted to Bob, while in the protocols of KA, both Alice and Bob
contribute information that is subsequently used to derive the shared
secret key. }Further, in a good KAS each party contributes equally
to the shared key and a dishonest party or a group of dishonest parties
cannot control or completely decide the final key. The last point
shows why all the traditional protocols of quantum cryptography e.g.,
BB84 \cite{bb84}, B92 \cite{b92}, ping-pong (PP) \cite{ping-pong},
LM05 \cite{lm05} etc. are not protocols of QKA in their original
forms.

Several protocols of classical key agreement are studied since the
well known Diffie-Hellman (DH) key agreement protocol or the exponential
key agreement protocol was introduced by Diffie and Hellman in 1976
\cite{Difiie-Hellman76}. A large number of the classical key agreement
protocols are actually variant of the DH protocol as they are based
on intractability of the DH problem {[}\cite{DF2} and references
therein{]}. To be precise, security of these protocols depends on
the intractability of discrete logarithm (DL) problem which may be
stated as follows: given a generator $g$ of a cyclic group $G$ and
an element $g^{x}$ in $G$, determine $x$. Quite similarly, the
DH problem is stated as: given $g^{x}$ and $g^{y}$, determine $g^{xy}$
\cite{DF3}. Clearly if we can solve DL problem in polynomial time
then we will be able to solve DH problem in polynomial time. As there
is no efficient classical algorithm for DL problem, modified and improved
DH protocols have been considered to be secure for long. Interestingly
in 1997, Shor introduced polynomial-time quantum algorithms for prime
factorization and discrete logarithms \cite{Shor}. These two quantum
algorithms clearly established that neither the RSA protocol nor the
DH based KA protocols would remain secure if a scalable quantum computer
is built. This fact along with the already established unconditional
security of QKD enhanced the interest on QKD and QKA. 

First protocol of QKA was introduced by Zhou et al. in 2004 \cite{N. Zhou (2004)}
using quantum teleportation. Almost simultaneously Hsueh and Chen
\cite{Hsuech and chen} proposed another protocol of QKA. However,
in 2009, Tsai and Hwang \cite{C.W. Tasi} showed that quantum teleportation
based Zhou et al. protocol was not a true protocol of QKA as a particular
user can completely determine the final (shared) key without being
detected. Next year Tsai et al. \cite{Tsai2} showed that even protocol
of Hsueh and Chen does not qualify as a protocol of QKA. In 2010,
Chong and Hwang \cite{QKA based on BB84} developed a protocol of
QKA using mutually unbiased bases (MUBs). Apparently Chong Hwang (CH)
protocol was the first successful protocol of QKA. They claimed that
their protocol is based on BB84. However, a deeper analysis would
show that their protocol is closer to LM05 protocol \cite{lm05}.
Of course the security of both LM05 and BB84 protocols arises from
the non-commutativity and nocloning principles. In 2011, Chong, Tsai
and Hwang \cite{Improvement on QKA} proposed a modified version of
Hsueh and Chen protocol that is free from the limitations of the original
protocol mentioned in Ref. \cite{Tsai2}. All the successful and unsuccessful
efforts of designing protocols of QKA until recent past were limited
to two party case. Recently an enhanced interest on multi-party QKA
schemes has been observed and several protocols have been reported
\cite{Three party QKA,MQKA,Improvements on MQKA,multi-party,MQKA-crypto analysis}.
A systematic review of all these existing works leads us to the following
observations. 
\begin{enumerate}
\item The amount of works reported to date on QKA is much less compared
to the amount of works reported on other aspects of quantum cryptography,
such as QKD, deterministic secure quantum communication (DSQC), quantum
secure direct communication (QSDC) and quantum dialogue (QD). Thus
we may conclude that QKA is not yet studied rigorously and probably
many more combinations of quantum states and protocols of QKA can
be found. Keeping this in mind we show that majority of the existing
protocols of QSDC, DSQC and QD can be turned into protocol of QKA
by introducing a delayed measurement technique.
\item Security of all the protocols of two-party and multi-party QKA reported
to date is based on conjugate coding, i.e., the security is obtained
using two or more MUBs and thus the protocols are essentially of BB84
type. This lead to a question: Is it essential to use non-orthogonal
states (2 or more MUBs) for designing of protocols of QKA? The question
is not yet answered, but the expected answer is {}``no'' as QKA
is related to QKD and a few orthogonal-states-based protocols of QKD
(e.g., Goldenberg-Vaidman (GV) protocol \cite{vaidman-goldenberg}
and N09 or counter-factual protocol \cite{N09}) are known since a
few years. Further, some of the present authors have recently shown
that protocols of QSDC and DSQC can be designed using orthogonal states
\cite{beyond-gv,preeti-arxiv}. In addition several exciting experiments
on orthogonal-state-based QKD are reported in recent past \cite{GV-experiment,N09-expt-1,N09 expt-2,N09 expt-3}.
These recent experimental observations and the recently proposed orthogonal-state-based
protocols are very interesting as they are fundamentally different
from the traditional conjugate coding based protocols where two or
more MUBs (set of non-orthogonal states) are used to provide security.
Keeping these in mind present paper aims to provide orthogonal-state-based
protocols of 2-party and multi-party QKA.
\end{enumerate}
Remaining part of the paper is organized as follows. In the next section
we present a protocol of QKA for 2-party scenario. In Section \ref{sec:Protocol-2-[Multiparty},
we provide a protocol of 3-party QKA and discuss the possibilities
of extending it to $n$-party ($n>3)$ scenario. Specifically, we
have shown that the proposed 3-party protocol can be extended to a
5-party protocol of QKA that uses 4-qubit $|\Omega\rangle$ state
or 4-qubit cluster state. In Section \ref{sec:Security-and-efficiency},
security and efficiency of the proposed protocols are discussed and
are compared with that of other existing protocols of QKA. In Section
\ref{sec:Turning-exisiting-protocols} we investigate the possibilities
of transforming the existing protocols of QSDC, DSQC and QD to protocols
of QKA. Finally the paper is concluded in Section \ref{sec:Conclusions}.

\section{Protocol 1: A 2-party orthogonal-state-based protocol of QKA\label{sec:Protocol-1} }
\begin{description}
\item [{Step~1:}] Alice prepares $|\psi^{+}\rangle^{\otimes n}$ where
$|\psi^{+}\rangle=\frac{|00\rangle+|11\rangle}{\sqrt{2}}$. She uses
first qubits of each Bell state to form an ordered sequences $p_{A}=\{p_{A}^{1},\, p_{A}^{2},\, p_{A}^{3},\cdots,\, p_{A}^{n}\}.$
Similarly she forms an ordered sequence $q_{A}=\{q_{A}^{1},\, q_{A}^{2},\, q{}_{A}^{3},\cdots,\, q_{A}^{n}\}$
with all the second qubits. Here $p_{A}^{i}$, $q_{A}^{i}$ denote
the first and second particles of $i^{th}$ copy of the Bell state
$|\psi^{+}\rangle,$ for $1\leq i\leq n$. She also prepares a random
sequence $K_{A}=\{K_{A}^{1},\, K_{A}^{2},\, K_{A}^{3},\cdots,\, K_{A}^{n}\}$,
where $K_{A}^{i}$ denotes the $i^{th}$ bit of sequence $K_{A}$
and $K_{A}^{i}$ is randomly chosen from $\{0,1\}.$ $K_{A}$ may
be considered as Alice's key.
\item [{Step~2:}] Alice prepares a sequence of $\frac{n}{2}$ Bell states
( $|\psi^{+}\rangle^{\otimes\frac{n}{2}}$) as decoy qubits and concatenates
the sequence with $q_{A}$ to form an extended sequence $q_{A}^{\prime}$.
She applies a permutation operator $\Pi_{2n}$ on $q_{A}^{\prime}$
to create a new sequence $\Pi_{2n}q_{A}^{\prime}=q_{A}^{\prime\prime}$
and sends that to Bob.
\item [{Step~3:}] After receiving the authentic acknowledgment of the
receipt of the entire sequence $q_{A}^{\prime\prime}$ from Bob, Alice
announces the coordinates of the qubits $(\Pi_{2n}$) sent by her.
Using the information Bob rearranges the qubits and performs Bell
measurements on the decoy qubits and computes the error rate. Ideally
in absence of Eve all the decoy Bell states are to be found in $|\psi^{+}\rangle.$
If the error rate is found to be within the tolerable limit, they
continue to the next step, otherwise they discard the protocol and
go back to \textbf{Step 1}.
\item [{Step~4:}] Bob drops the decoy qubits to obtain $q_{A}$. Now he
prepares a new random sequence $K_{B}=\{K_{B}^{1},\, K_{B}^{2},\, K_{B}^{3},\cdots,\, K_{B}^{n}\}$,
where $K_{B}^{i}$ denote the $i^{th}$ bit of sequence $K_{B}$,
 for $1\leq i\leq n$ and $K_{B}^{i}$ is randomly chosen from $\{0,1\}.$
$K_{B}$ may be considered as Bob's key. He applies a unitary operation
on each qubit of sequence $q_{A}$ to encode $K_{B}$. The encoding
scheme is as follows: to encode $K_{B}^{i}=0$ and $K_{B}^{i}=1$
he applies $I$ and $X$ respectively on $q_{A}^{i}$. This forms
a new sequence $q_{B}$. After encoding operation, Bob concatenates
$q_{B}$ with a sequence of $\frac{n}{2}$ Bell states ( $|\psi^{+}\rangle^{\otimes\frac{n}{2}}$)
that is prepared as decoy qubits and subsequently applies the permutation
operator $\Pi_{2n}$ to obtain an extended and randomized sequence
$q_{B}^{\prime}$ which he sends to Alice. 
\item [{Step~5:}] After receiving the authenticated acknowledgment of
the receipt of the entire sequence $q_{B}^{\prime}$ from Alice, Bob
announces the position of the decoy qubits (note that he does not
disclose the actual order of the message qubits) i.e., $\Pi_{n}\in\Pi_{2n}.$
Alice checks the possibility of eavesdropping by following the same
procedure as in \textbf{Step 3}. If the error rate is found to be
within the tolerable limit, they continue to the next step, otherwise
they discard the protocol and go back to \textbf{Step 1}. 
\item [{Step~6:}] Alice publicly announces her key $K_{A}$ and Bob uses
that and his own key (sequence) $K_{B}$ to form the shared key: $K=K_{A}\oplus K_{B}$. 
\item [{Step~7:}] Bob announces the actual order of the message qubits
$(\Pi_{n}\in\Pi_{2n})$ and Alice uses that information to obtain
$q_{B}.$ Now she combines $p_{A}$ and $q_{B}$ and performs Bell
measurements on $p_{A}^{i}q_{B}^{i}$. This would reveal $K_{B}$
as she knows the initial state and the encoding scheme used by Bob. 
\item [{Step~8:}] Using $K_{A}$ and $K_{B}$ Alice prepares her copy
of the shared key i.e., $K=K_{A}\oplus K_{B}$. 
\end{description}
The protocol discussed above is an orthogonal-state-based 2-party
protocol of QKA. However, several multi-party protocols of QKA are
introduced in recent past \cite{Three party QKA,MQKA,Improvements on MQKA,multi-party,MQKA-crypto analysis}.
Of course none of these recently introduced multi-party QKA protocols
are based on orthogonal state. Keeping these in mind we aim to provide
a completely orthogonal-state-based protocol of 3-party QKA along
the line of \cite{Three party QKA}. Further the possibility of extending
the proposed orthogonal-state-based three-party protocol into $n$-party
case with $n>3$ is also discussed in the following section.

\section{Protocol 2: A multi-party protocol of QKA\label{sec:Protocol-2-[Multiparty}}

In analogy to the previous protocol Alice, Bob and Charlie produce
their secret keys: 

\[
\begin{array}{lc}
K_{A} & =\{K_{A}^{1},\, K_{A}^{2},\, K_{A}^{3},\cdots,\, K_{A}^{n}\},\\
K_{B} & =\{K_{B}^{1},\, K_{B}^{2},\, K_{B}^{3},\cdots,\, K_{B}^{n}\},\\
K_{C} & =\{K_{C}^{1},\, K_{C}^{2},\, K_{C}^{3},\cdots,\, K_{C}^{n}\},\end{array}\]
where $K_{A}^{i},\, K_{B}^{i},\, K_{C}^{i}$ denote $i^{th}$ bit
of key of Alice, Bob and Charlie respectively%
\footnote{Here subscripts A, B, C denote Alice, Bob and Charlie, respectively. %
} and $i=1,\,2,\cdots,\, n$. We describe a protocol of multi-party
QKA in the following steps. 
\begin{description}
\item [{Step~1:}] Alice, Bob and Charlie separately prepare $|\psi^{+}\rangle_{A}^{\otimes n}$,
$|\psi^{+}\rangle_{B}^{\otimes n}$ and $|\psi^{+}\rangle_{C}^{\otimes n}$,
respectively. As in \textbf{Step 1} of the previous protocol Alice
prepares two ordered sequences $p_{A}=\{p_{A}^{1},\, p_{A}^{2},\, p_{A}^{3},\cdots,\, p_{A}^{n}\}$
and $q_{A}=\{q_{A}^{1},\, q_{A}^{2},\, q_{A}^{3},\cdots,\, q_{A}^{n}\}$
composed of all the first and the second qubits of the Bell states
that she has prepared. Similarly, Bob and Charlie prepare $p_{B}=\{p_{B}^{1},\, p_{B}^{2},\, p_{B}^{3},\cdots,\, p_{B}^{n}\}$,
$q_{B}=\{q_{B}^{1},\, q_{B}^{2},\, q_{B}^{3},\cdots,\, q_{B}^{n}\}$
and $p_{C}=\{p_{C}^{1},\, p_{C}^{2},\, p_{C}^{3},\cdots,\, p_{C}^{n}\}$,
$q_{C}=\{q_{C}^{1},\, q_{C}^{2},\, q_{C}^{3},\cdots,\, q_{C}^{n}\}$
from $|\psi^{+}\rangle_{B}^{\otimes n}$ and $|\psi^{+}\rangle_{C}^{\otimes n}$,
respectively. 
\item [{Step~2:}] Each of Alice, Bob and Charlie separately prepares sequence
of $\frac{n}{2}$ Bell states ( $|\psi^{+}\rangle^{\otimes\frac{n}{2}})_{j}$
with $j\in\{A,\, B,\, C\}$ as decoy qubits and concatenates the sequence
with $q_{j}$ to form extended sequences $q_{j}^{\prime}$. Subsequently
user $j$ applies permutation operator $(\Pi_{2n})_{j}$ on $q_{j}^{\prime}$
to create a new sequence $(\Pi_{n})_{j}q_{j}^{\prime}=q_{j}^{\prime\prime}$
and sends that to user $j+1$.\\
Here we follow a notation in which $j\in\{A,\, B,\, C\}$ and
$A,\, B,C$ follows a modulo $3$ algebra that gives us the relations:
$A+3=B+2=C+1=A,$ $A=C+1,\, B=A+1,\,$ $C=B+1$ and so on.
\item [{Step~3:}] After receiving the authentic acknowledgment of receipt
from the receiver (user $j+1$) corresponding sender (user $j$) announces
the coordinates of the qubits $(\Pi_{2n})_{j}$ sent by him/her. Each
receiver computes error rate as in \textbf{Step 3} of the previous
protocol. If the computed error rates are found to be within the tolerable
limit, they continue to the next step, otherwise they discard the
protocol and go back to \textbf{Step 1}.
\item [{Step~4:}] After discarding the decoy qubits each user $j$ encodes
his/her secret bits by applying the unitary operation on each qubit
of the sequence received by him (i.e., on $q_{j-1}$) in accordance
with his/her key $K_{j}$. The encoding scheme is as follows: If $K_{j}^{i}=0\,(1)$
then user $j$ applies $I\,(X)$ on $q_{j-1}^{i}.$ As a result of
encoding operations, user $j$ obtains a new sequence $r_{j}$. After
the encoding operation user $j$ concatenates $r_{j}$ with a sequence
of $\frac{n}{2}$ Bell states ( $|\psi^{+}\rangle^{\otimes\frac{n}{2}})_{j}$
that is prepared as decoy qubits and subsequently applies the permutation
operator $(\Pi_{2n})_{j}$ to obtain an extended and randomized sequence
$r_{j}^{\prime}$ which he/she sends to the user $j+1$. 
\item [{Step~5:}] After receiving the authentic acknowledgment of the
receipt of the sequence $r_{j}^{\prime}$ from the receiver $j+1$,
the sender $j$ announces the coordinates of the decoy qubits i.e.,
$(\Pi_{n})_{j}\in(\Pi_{2n})_{j}$. User $j+1$ uses the information
for computing the error rate as before and if it is below the threshold
value then they go on to the next step, otherwise they discard the
communication. In absence of eavesdropping user $j$ announces the
coordinates of the message qubits i.e., $(\Pi_{n})_{j}\in(\Pi_{2n})_{j}$. 
\item [{Step~6:}] Same as \textbf{Step 4} with only difference that if
$K_{j}^{i}=0$ and $K_{j}^{i}=1$ then user $j$ applies $I$ and
$Z$ respectively on $r_{j-1}^{i}.$ As a result of encoding operations
user $j$ obtains a new sequence $s_{j}$ and after insertion of decoy
qubits and applying permutation operator he/she obtains a randomized
sequence $s_{j}^{\prime}$ which he/she sends to the user $j+1$.
\item [{Step~7:}] Same as \textbf{Step 5}.
\item [{Step~8:}] After discarding the decoy qubits each user rearranges
the sequence received by him/her. Now each user $j$ has two ordered
sequences $p_{j}$ and $s_{j-1}.$ Each of the users $j$ performs
Bell measurements on $p_{j}^{i}s_{j-1}^{i}$. According to the output
of the Bell measurement and Table \ref{tab:Table} each user $j$
can obtain the secret keys of the other two parties. Hence the shared
secret key $K=K_{A}\oplus K_{B}\oplus K_{C}$ can be generated.
\end{description}
\begin{table}[H]
\begin{centering}
\begin{tabular}{|>{\centering}p{1in}|>{\centering}p{1in}|>{\centering}p{1in}|c|}
\hline 
Initial state prepared by user $j$ & First operator applied by user $j+1$ & Second operator applied by user $j+2$ & Final State\tabularnewline
\hline
$|\psi^{+}\rangle$ & $I\otimes I$ & $I\otimes I$ & $|\psi^{+}\rangle$\tabularnewline
\cline{2-4} 
 & $I\otimes I$ & $I\otimes Z$ & $|\psi^{-}\rangle$\tabularnewline
\cline{2-4} 
 & $I\otimes X$ & $I\otimes I$ & $|\phi^{+}\rangle$\tabularnewline
\cline{2-4} 
 & $I\otimes X$ & $I\otimes Z$ & $|\phi^{-}\rangle$\tabularnewline
\hline
\end{tabular}
\par\end{centering}

\caption{\label{tab:Table}Transformation of $|\psi^{+}\rangle$ based on two
operations. Here $+$ refers to modulo 3 operations. $j\in\{A,\, B,\, C\}$
where $A,\, B,\, C$ stands for Alice, Bob and Charlie, respectively.
Thus $A+2=C=A-1$ and so on. Further, to denote the Bell states, we
have used the following conventions: $|\psi^{\pm}\rangle=\frac{1}{\sqrt{2}}(|00\rangle\pm|11\rangle)$
and $|\phi^{\pm}\rangle=\frac{1}{\sqrt{2}}(|01\rangle\pm|10\rangle)$.}
\end{table}

Here we note that $\{I,\, X,\, iY,\, Z\}$ is a modified Pauli group%
\footnote{In the stabilizer formalism of quantum error correction Pauli group
is frequently used (see Section 10.5.1 of \cite{Nielsenchuang}).
It is usually defined as $G_{1}=\left\{ \pm I,\pm iI,\pm\sigma_{x},\pm i\sigma_{x},\pm\sigma_{y},\pm i\sigma_{y},\pm\sigma_{z},\pm i\sigma_{z}\right\} ,$
where $\sigma_{i}$ is a Pauli matrix. The inclusion of $\pm1$ and
$\pm i$ ensures that $G_{1}$ is closed under standard matrix multiplication,
but the effect of $\sigma_{i},\,-\sigma_{i},\, i\sigma_{i}$ and $-i\sigma_{i}$
on a quantum state is the same. So in \cite{qd} we redefined the
multiplication operation for two elements of the group in such a way
that global phase is ignored from the product of matrices. This is
consistent with the quantum mechanics and it gives us a modified Pauli
group $G_{1}=\{I,\,\sigma_{x},\, i\sigma_{y},\,\sigma_{z}\}=\{I,\, X,\, iY,\, Z\}.$%
} under multiplication and $\{I,\, X\},$ $\{I,\, Z\}$ are its disjoint
subgroups. Here disjoint subgroups refer to two subgroups $g_{i}$
and $g_{j}$ of a group $G$ that satisfy $g_{i}\cap g_{j}=\{I\},$
where $I$ is the identity element. Thus except identity element $g_{i}$
and $g_{j}$ do not contain any other common element. Now we assume
that $G$ is a group of order $M$ under multiplication and elements
of $G$ are $x$-qubit unitary operators. Further, we assume that
there exist $n$ mutually disjoint subgroups $g_{i}$ with $i=1,\cdots,\, n$
of the group $G$ such that $g_{i}$'s are of equal size (say each
of the $g_{i}$'s has $2^{y}$ elements) and $\Pi_{i}^{\otimes m}g_{i}=g_{1}\otimes g_{2}\otimes g_{3}\otimes\cdots\otimes g_{m}=\left\{ U_{1},U_{2},\cdots,U_{(2^{y})^{m}}\right\} $
where $\left(2^{y}\right)^{m}\leq M$; $U_{i}\in G$ and $U_{i}\neq U_{l}$
$\forall\, i,\, l\in\left\{ 1,2,\cdots,\left(2^{y}\right)^{m}\right\} .$
Now if we have $I^{\otimes\left(N-x\right)}U_{i}|\phi_{0}\rangle=|\phi_{i}\rangle$
and $\langle\phi_{i}|\phi_{l}\rangle=\delta_{i,l}$ where $|\phi_{i}\rangle$
is an $N$-qubit quantum state with $N>x,$ then we can have an $(m+1)$-party
version of Protocol 2 of QKA. In this $(m+1)$-party protocol of QKA
all the $(m+1)$ parties create quantum state $|\phi_{0}\rangle$
in the beginning. Each user keeps the first $N-x$ qubits of $|\phi_{0}\rangle$
with himself/herself and sends the remaining qubits to the user $j+1$
after following the strategy for eavesdropping checking. Subsequently,
user $j$ encodes his/her $y$-bit secret key ($N>x\geq y$) by applying
unitary operators from $g_{1}$ on the $x$ qubits that he/she has
received from the user $j-1$ in the previous step and sends the key
encoded state to user $j+1$. After $m$ rounds of such encoding (in
$k^{{\rm th}}$ round of encoding operation all the users encodes
their keys using elements of $g_{k}$) and communication operations
user $j$ measures the $N$ qubits of his/her possession using $\{|\phi_{i}\rangle\}$
basis. From the input state ($|\phi_{0}\rangle$) and output state
(say, $|\phi_{{\rm {\rm final}}}\rangle=|\phi_{k}\rangle)$ he/she
would know the unitary operator $U_{k}$ that has converted the initial
state into the final state. Now the condition $\Pi_{i}^{\otimes m}g_{i}=\left\{ U_{1},U_{2},\cdots,U_{(2^{y})^{m}}\right\} $
where $U_{i}\in G$ and $U_{i}\neq U_{l}$ ensures that every sequence
of encoding operations will lead to different $U_{k}$ and this is
how user $j$ can know the key encoded by the other users and he/she
can use that to create the shared key $K_{1}\oplus K_{2}\oplus\cdots\oplus K_{m},$
where the secret key of the user $j$ is $K_{j}.$ 

In Protocol 2 we have used modified Pauli group $G=G_{1}=\{I,\, X,\, iY,\, Z\}.$
It has three disjoint subgroups: $g_{1}=\{I,X\},\, g_{2}=\{I,Z\},\, g_{3}=\{I,iY\}$
which satisfy $g_{1}\otimes g_{2}=g_{2}\otimes g_{3}=g_{3}\otimes g_{1}=G_{1}.$
Further, $|\phi_{0}\rangle=|\psi^{+}\rangle$ and as $G_{1}$ is the
set of elements used for dense coding using Bell states so it naturally
implies $U_{i}|\phi_{0}\rangle=|\phi_{i}\rangle\forall\, U_{i}\in G:$
$\langle\phi_{i}|\phi_{l}\rangle=\delta_{i,l}$. Thus Protocol 2 is
a special case of a more general scenario described here. Many more
examples can be obtained from the properties of Pauli groups discussed
in Ref. \cite{qd}. Just to provide specific examples we may note
that for the modified Pauli group \begin{equation}
\begin{array}{lcl}
G_{2} & = & G_{1}\otimes G_{1}=\{I,\, X,\, iY,\, Z\}\otimes\{I,\, X,\, iY,\, Z\}\\
 & = & \left\{ I\otimes I,\, I\otimes X,\, I\otimes iY,\, I\otimes Z,\, X\otimes I,\, X\otimes X,\right.\\
 &  & X\otimes iY,\, X\otimes Z,\, iY\otimes I,\, iY\otimes X,\, iY\otimes iY,\\
 &  & \left.iY\otimes Z,Z\otimes I,\, Z\otimes X,\, Z\otimes iY,\, Z\otimes Z\right\} \end{array}\label{eq:pustak2}\end{equation}
we have following disjoint subgroups of order 2: $g_{1}=\left\{ I\otimes I,\, I\otimes X\right\} ,$
$g_{2}=\left\{ I\otimes I,\, X\otimes I\right\} ,$ $g_{3}=\left\{ I\otimes I,\, I\otimes Z\right\} ,$
$g_{4}=\left\{ I\otimes I,\, Z\otimes I\right\} ,$ $g_{5}=\left\{ I\otimes I,\, I\otimes iY\right\} $
and $g_{6}=\left\{ I\otimes I,\, iY\otimes I\right\} .$ Further,
these disjoint subgroups satisfy \begin{equation}
g_{1}\otimes g_{2}\otimes g_{3}\otimes g_{4}=g_{1}\otimes g_{2}\otimes g_{5}\otimes g_{6}=g_{3}\otimes g_{4}\otimes g_{5}\otimes g_{6}=G_{2}\label{eq:pustak3}\end{equation}
and the elements of $G_{2}$ can be used for dense coding using 4-qubit
maximally entangled $|\Omega\rangle$ state and cluster ($|C\rangle)$
state if the elements of $G_{2}$ operate on $1^{{\rm st}}$ and $3^{{\rm rd}}$
qubits of these states. Here \[
\begin{array}{lcl}
|\Omega\rangle & = & \frac{1}{2}(|0000\rangle+|0110\rangle+|1001\rangle-|1111\rangle),\\
|C\rangle & = & \frac{1}{2}(|0000\rangle+|0011\rangle+|1100\rangle-|1111\rangle).\end{array}\]
The table of dense coding for these states using elements of $G_{2}$
is explicitly shown in our earlier work (see Table 1 of Ref. \cite{qd}).
As the elements of $G_{2}$ can be used for dense coding using $|\Omega\rangle$
and $|C\rangle$ states, output states obtained on application of
the elements of $G_{2}$ on $|\Omega\rangle$ or $|C\rangle$ are
mutually orthogonal. This clearly implies that we can construct a
5-party protocol of QKA using $|\Omega\rangle$ or $|C\rangle$ state
where each user prepares a large number copies of one of these two
states and keeps 2nd and 4th qubit with himself/herself and sends
the remaining qubits to next user and later encodes his/her secret
key using $g_{i}'$s. In a 5-party protocol, encoding operation should
take place in 4 rounds and the users would use either $g_{1},g_{2},g_{3},g_{4}$
or $g_{1},g_{2},g_{5},g_{6}$ or $g_{3},g_{4},g_{5},g_{6}.$ We can
generate many more examples of multi-party protocols of QKA using
similar strategy and properties of modified Pauli group.

\section{Security and efficiency analysis \label{sec:Security-and-efficiency}}

Protocol 2 is designed along the line of existing protocol of Yin,
Ma and Liu \cite{Three party QKA} with a modified strategy of eavesdropping
checking that converts the non-orthogonal-state-based protocol of
Yin, Ma and Liu into an orthogonal-state-based protocol. Unconditional
security of the eavesdropping checking using this technique is already
shown in our earlier works \cite{beyond-gv,preeti-arxiv} where we
have also established that security of this orthogonal-state-based
technique of eavesdropping checking originates from the monogamy of
entanglement \cite{preeti-arxiv}. Thus the protocol is secure against
external attacks (eavesdropping). Remaining part of the protocol is
technically equivalent to Yin Ma Liu (YML) protocol and consequently
the security of YML protocol against the internal attacks (i.e., the
attempts of malicious Alice, Bob and Charlie to completely control
the key either individually or by mutual cooperation of any two users)
is applicable here, too. Thus Protocol 2 is a secure protocol of QKA
and it does not need any separate elaborate discussion. Keeping this
in mind in the remaining part of the present section we have explicitly
analyzed the security of Protocol 1.

\subsection{Security against eavesdropping}

Our Protocol 1 and also the protocol of Chong and Hwang \cite{QKA based on BB84}
may be viewed as protocols of secure direct communication of $K_{B}$
from Bob to Alice added with a classical communication of $K_{A}$
from Alice to Bob. Specifically, instead of sending a meaningful message
Alice and Bob send random keys to each other. While Bob sends his
key $K_{B}$ by using a DSQC or QSDC scheme, Alice announces her key
$K_{A}$ publicly. Security proofs of the existing protocols of DSQC
and QSDC ensures that the key communicated by Bob (i.e., $K_{B})$
using DSQC or QSDC scheme is unconditionally secure. Thus Eve has
no information about $K_{B}.$ On the other hand the key communicated
by Alice (i.e., $K_{A}$) is a public knowledge. However, it does
not affect the secrecy of the shared key as the final shared key to
be produced and used is $K_{A}\oplus K_{B},$ knowledge of $K_{A}$
alone does not provide any information about $K_{A}\oplus K_{B}$.
Thus the shared key produced in this manner is secure from external
attacks of Eve. However, there may exist insider attacks in which
Alice or Bob tries to completely control the shared key. Security
of Protocol 1 against such attacks is described below.

\subsubsection{Security against dishonest Alice}

To communicate $K_{B}$ if Alica and Bob use a standard protocol of
DSQC or QSDC (say they use PP protocol), then it would be possible
for Alice to know Bob's secret key before she announces $K_{A}.$
In that case she will be able to completely control the shared key
by manipulating $K_{A}$ as per her wish. To circumvent this attack
we have modified the protocol in such a way that Bob does not announce
the coordinates of the message qubits sent by him till he receives
$K_{A}.$ This strategy introduces a delay in measurement of Alice
and this delayed measurement strategy ensures that Alice cannot control
the key by knowing $K_{B}$ prior to her announcement of $K_{A}$.

\subsubsection{Security against dishonest Bob}

Alice announces her key only after receiving the message qubits (without
their actual order) from Bob. This ensures that Bob cannot control
the key by knowing Alice's key. Only thing that Bob can do after knowing
$K_{A}$ is to change/modify the coordinates of $q_{B}^{\prime}$,
but any modification in that would lead to entanglement swapping in
our case and that would lead to probabilistic outcomes without any
control of Bob. Further, Bob will be completely unaware of $K_{B}$
to be generated by Alice in that case and as a consequence any such
effort of Bob would lead to different keys at ends of Alice and Bob.
Thus the protocol ensures that Bob cannot control the key. Here we
may note that similar strategy was used in Chong and Hwang \cite{QKA based on BB84}
protocol. In their protocol modified QSDC scheme that was used for
Bob to Alice communication was equivalent to LM05 \cite{lm05} protocol.
In contrast here we have used a modified orthogonal version of PP-type
protocol which may be referred to as ${\rm PP^{GV}}$ protocol \cite{preeti-arxiv}.

\section{Turning existing protocols of quantum communication to protocols
of QKA\label{sec:Turning-exisiting-protocols}}

In the previous sections we have seen that there exist a strong link
between protocols of DSQC/QSDC and those of QKA. For example, PP \cite{ping-pong}
and LM05 \cite{lm05} protocols of QSDC have already been employed
to design protocols of QKA (Protocol 1 presented here and CH protocol
\cite{QKA based on BB84}). This observation leads to an important
question: Is it possible to convert all protocols of secure direct
quantum communication into protocols of QKA? In what follows we aim
to answer this question. We also aim to study the possibilities of
transforming other protocols of quantum communication to protocols
of QKA.

\subsection{Turning a protocol of QSDC/DSQC to a protocol of QKA}

Recently we have shown that maximally efficient protocols for secure
direct quantum communications can be constructed using any arbitrary
orthogonal basis \cite{beyond-gv}. However, all of them will not
lead to protocol of QKA. To be precise, eavesdropping can be avoided
in all protocols of DSQC and QSDC and by randomizing the sequence
of key encoded bits sent by Bob (i.e., by delaying the measurement
to be performed by Alice) we can circumvent the attacks of dishonest
Alice, but it is not sufficient to build a protocol of QKA. We also
need to avoid the attacks of dishonest Bob. To do so we need to restrict
the information available to Bob. Specifically, Bob must not have
complete information of the basis that is used to prepare the qubits
on in which he has encoded his key. In our Protocol 1 and in all orthogonal-state-based
two-way DSQC/QSDC protocols this can be achieved if Alice keeps some
of the qubits of each entangled state with her as that would restrict
Bob from changing $K_{B}$ after receiving $K_{A}$. The same can
be achieved in a non-orthogonal-state-based protocol by using more
than one MUBs. If Alice prepares the state randomly using one of the
basis sets and don't disclose the basis set used by her till Bob discloses
the sequence then Bob will not have complete access of the basis set
used for preparation of the message qubits. As a consequence he will
not be able to control the key. This is shown in a particular case
in Ref. \cite{QKA based on BB84}.

The above discussion shows that the DSQC/QSDC protocol to be used
to implement a QKA protocol cannot be one-way as in that case Bob
will have complete access to the basis in which the quantum state
used for encoding of his key is prepared (since in a one-way protocol
Bob himself will prepare the quantum state). Thus none of the one-way
protocol of DSQC or QSDC would lead to QKA. However, most of the two-way
protocols of secure quantum communication would lead to QKA. As example,
we may note both Deng Long Liu (DLL) protocol \cite{DLL} and Cai
Li (CL) protocol \cite{CL} can be viewed as variant of PP protocol
\cite{my book}, but DLL being a one-way protocol would not give us
a QKA protocol, but two-way CL protocol would lead to a QKA protocol.

\subsection{Turning a protocol of QD to a protocol of QKA}

A very interesting two-way quantum communication scheme is QD {[}\cite{qd}
and references therein{]}. Since in the above we have already seen
that two-way secure direct communication is useful for QKA and since
a large number of alternatives for implementing quantum dialogue are
recently proposed by us (see Table 4 of Ref. \cite{qd}), it would
be worthy to investigate the relation between QKA and QD. In a Ba
An type QD protocol \cite{ba-an}, Alice keeps part of an entangled
state ($|\phi\rangle_{i})$ with herself and encodes her secret on
the remaining qubits by applying unitary operation $U_{A}$ and subsequently
sends the message encoded qubits to Bob who applies $U_{B}$ on them
and returns the qubits to Alice with appropriate strategy of eavesdropping
checking. Now Alice measures the final state ($|\phi\rangle_{f})$
and announces the outcome. As the states and operators are chosen
in such a way that $|\phi\rangle_{i}$ and $|\phi\rangle_{f}$ are
mutually orthogonal, from the announcement of Alice we know $U_{A}U_{B}$.
As Alice (Bob) knows $U_{A}\,(U_{B})$ she (he) can easily obtain
$U_{B}\,(U_{A})$ using $U_{A}U_{B}$ obtained from the announcement
of Alice. For a detailed discussion see Ref. \cite{qd} where it is
explicitly shown that if we have a set of mutually orthogonal $n$-qubit
states $\{|\phi_{0}\rangle,|\phi_{1}\rangle,\cdots,|\phi_{i}\rangle,\cdots,|\phi_{2^{n}-1}\rangle\}$
and a set of $m$-qubit unitary operators $\{U_{0},U_{1},U_{2},\cdots,U_{2^{n}-1}\}$
such that $U_{i}|\phi_{0}\rangle=|\phi_{i}\rangle$ and $\{U_{0},U_{1},U_{2},\cdots,U_{2^{n}-1}\}$
forms a group under multiplication then it would be sufficient to
construct a quantum dialogue protocol of Ba An type. Now assume that
$n>m$ and Alice encodes nothing (i.e., she always choose $U_{A}=I_{m}$
) and keeps $(n-m)$-qubits with herself and sends the remaining $m$-qubits
to Bob who encodes his key by applying an $m$-qubit unitary operation
$U_{B}$ and sends that back to Alice, but only after changing the
order so that Alice cannot measure the final state immediately. Alice
announces her key after receiving the key encoded qubits from Bob
as in Protocol 1 and subsequently Bob announces the sequence of the
message qubits sent by him. In QKA Alice does not need to disclose
her measurement outcome. This modified QD protocol is equivalent to
our Protocol 1. This clearly shows that all protocols of QD with $n>m$
would lead to protocols of QKA. It is interesting because in \cite{qd}
we have shown that a large number of alternative combinations of quantum
states and unitary operators can be used to implement QD. All of them
(if $n>m$) will be useful for QKA, too.

\subsection{Efficiency analysis}

A well-known measure of efficiency of secure quantum communication
is known as qubit efficiency \cite{defn      of      qubit      efficiency}
which is given as \begin{equation}
\eta=\frac{c}{q+b},\label{eq:efficiency
    2}\end{equation}
where $c$ denotes the total number of transmitted classical bits
(message bits), $q$ denotes the total number of qubits used and $b$
is the number of classical bits exchanged for decoding of the message
(classical communication used for checking of eavesdropping is not
counted). This measure was introduced by Cabello in 2000 and it has
been frequently used since then to compare protocols of secure direct
communication. As we are not interested in communicating a message
here, so we may modify the meaning of $c$ in $\eta_{2}$ to make
it suitable for comparison of protocols of QKA. In the modified notion
$c$ is the length of the shared key generated by the protocol. Thus
in case of our first protocol if we generate an $n$-bit shared key
then $c=n.$ Further, in the entire protocol we have used $2n$ Bell
states i.e., $4n$ qubits (of which $n$-Bell states were used as
decoy qubits). Thus $q=4n$. Now Alice and Bob announces the coordinates
of the message qubits and Alice announces $K_{A}$, each of these
three steps require communication of $n$ classical bits. Thus $b=3n.$
All other classical communications incurred in the process are related
to the checking of eavesdropping and classical bits exchanged for
eavesdropping checking are not counted in $b$. Thus $b=3n.$ This
makes $\eta=\frac{n}{4n+3n}=\frac{1}{7}=14.29\%.$ In the similar
manner if an $n$-bit shared key is prepared through Protocol 2 then
$c=n$ and $q=3(2n+3n)$ as each party creates $n$ Bell states for
key encryption and $\frac{3n}{2}$ Bell states for eavesdropping checking.
Further, each party uses $3n$ bits of classical information for the
disclosure of coordinates of the message qubits. Thus $b=3\times3n=9n$
and consequently $\eta=\frac{n}{15n+9n}=\frac{1}{24}=4.17\%.$ As
YML protocol is similar to the Protocol 2 with only difference in
the strategy adopted for eavesdropping checking, for YML protocol
also we obtain $\eta=4.17\%$. Clearly, Protocol 1 is more efficient
than Protocol 2 and YML protocol, but Protocol 1 is less efficient
than its QSDC counterpart (${\rm PP^{GV}}$ protocol) whose qubit
efficiency as per the unmodified definition is $\eta=\frac{n}{4n+2n}=\frac{1}{6}=16.67\%$%
\footnote{In ${\rm PP^{GV}}$Alice does not need to disclose her key $K_{A}$.
Everything else is the same and as a consequence $b=2n,\, q=4n$ and
$c=n$ with $c$ being the number of bits in the message or key that
is transmitted.%
}. This is expected as with the increase on number of parties contributing
to the key, $q$ and $b$ required to generate the key of same size
should also increase. This point can be further established by noting
that $\eta$ for the 5-party protocol described above will be $\frac{1}{70}=1.43\%$
as $q=4n\times5=20n,$ $b=2n\times5\times5=50n$ and $c=n.$

\section{Conclusions\label{sec:Conclusions}}

In the present work we have proposed two orthogonal-state-based protocols
of QKA. The first one works for 2-party case and the second one works
for multi-party case. These are first set of orthogonal-state-based
protocols of QKA as all the existing protocols of QKA are based on
conjugate coding. Thus the proposed protocols are fundamentally different
from all the existing protocols of QKA. Orthogonal-state-based protocols
show that the use of conjugate coding or in other words use of non-commutativity
principle is not essentially required for unconditional security.
Thus it requires lesser quantum resources in a sense. To be precise,
monogamy of entanglement is sufficient to protect these protocols
\cite{preeti-arxiv}. We have also shown that most of the existing
protocols of QSDC and DSQC and all the protocols of QD can be turned
into protocols of QKA. Thus the present work leads to several new
options for implementation of QKA. Further, as the orthogonal-state-based
protocols of QSDC and QKD are experimentally implemented in recent
past, the protocols proposed here seem to be experimentally realizable. 

\textbf{Acknowledgment:} AP thanks Department of Science and Technology
(DST), India for support provided through the DST project No. SR/S2/LOP-0012/2010
and he also acknowledges the supports received from the projects CZ.1.05/2.1.00/03.0058
and CZ.1.07/2.3.00/20.0017 of the Ministry of Education, Youth and
Sports of the Czech Republic.

\end{document}